\documentstyle[preprint,aps,12pt]{revtex}
\begin{document}
\draft
\title{ Self-Similar Potentials and q-Coherent States}
\author{ Takahiro Fukui}
\address{Yukawa Institute for Theoretical Physics, Kyoto University,
kyoto 606-01, Japan}
\date{September 1993}
\maketitle

\begin{abstract} 
The self-similar potentails is formulated in terms of the
shape-invariance. Based on it, a coherent state associated with
the shape-invariant potentials is calculated in case of the
self-similar potentials. It is shown that it reduces to
the q-deformed coherent state.
\end{abstract}



The term ``coherent states" is applied to many objects today\cite{KL}.
The ordinary coherent state, which is defined as an eigenstate
of the boson anihilation operator, has a property of
the non-orthogonality and the over-completeness in $L^2$.
It is closely related to the irreducible representation
of the Heisenberg-Weyl group, and this property is generalized\cite{PE}
to a wide class of continuous groups with square-integrable
representation. Namely, the so-called generalized coherent
states are family of vectors $U(g)|0\rangle$, where $U$ is
a continuous irreducible representataion, $g$ is an element of
a group and $|0\rangle$ is a vector in the representaion space.
These coherent states also share in common the property of
non-orthogonality and over-completeness.

Recently, a ``coherent state" is proposed from a
quite different point of view\cite{FU}.
It is closely tied to the shape-invariance property of potentials.
In quantum mechanical one-body problems, a lot of
exactly solved potentials are known.
For a large class of such potentials, it is shown
that hamiltonians have a property of reparametrization invariance,
which is today called the shape-invariance\cite{IN,GE}.
For such potentials we can
obtain the eigenvalues and eigenstates by an algebraic way,
using parameter-dependent ``creation operators" and their
shape-invariance condition. An associated coherent state is defined as
an eigenstate of the ``annihilation" operator introduced when the
shape-invariance condition is represented as a commutation relation.

On the other hand, interesting solved potentials with a property of
self-similarity are proposed by Shabat\cite{SH} and Spiridonov\cite{SP}.
What is suprised is that they have the $su_q$(1,1)
dynamical symmetry\cite{SP,SK}.
These potentials are, therefore, one of the realization of q-deformed
algebras in physics,
and it is worth examining them furthermore in order to
consider a role which the q-deformation plays in physics.

The purpose of this letter is to examine an associated
coherent state introduced
in\cite{FU} in case of the self-similar potentials.
To achieve this, we first reformulate the self-similar potentials in
terms of the shape-invariance, and next calculte the coherent state
for these potentials.

Let us consider the following operators
\begin{equation}
D_n\equiv\frac{1}{\sqrt{2}}\left(W_n(x)+\frac{d}{dx}\right) ,
\end{equation}
where $W_n(x)$ is the so-called superpotential.
These and their hermitian conjugates are considered as generalized
creation and annihilation operators. We here introduce the relation
\begin{equation}
D_nD_n^\dagger =D_{n+1}^\dagger D_{n+1}+R_{n+1}\quad (n=0,1,2,\cdots ),
\end{equation}
where $R_{n+1}$ is a constant independent of $x$. In terms of
the superpotentials this relation is given by
\begin{equation}
W_n^2+W_n'=W_{n+1}^2-W_{n+1}'+2R_{n+1}\quad
(n=0,1,2,\cdots ),
\end{equation}
where dash denotes the derivative with respect to $x$.
This infinite chain of differential equations reduce to only one
equation if we adopt an anzats
\begin{equation}
W_n(x)=W(x,a_n), \quad
a_n\equiv\underbrace{f(f(\cdots f(}_{\mbox{$n$ times}}a_0)\cdots )),
\end{equation}
where $a_n$ is a parameter generated from $a_0$ by a function $f$.
In fact, substituting this into Eq.(3), we can easily confirm
the following statement: If the relation
\begin{equation}
W^2(x,a)+W'(x,a)=W^2(x,f(a))-W'(x,f(a))+2R(f(a))
\end{equation}
is satisfied identically with respect to $a$, the set of equations (3)
do not depend on $n$.
Now let us denote $ D_n\equiv D(a_n)$. Then we have the relation
\begin{equation}
D(a)D^\dagger (a)=D^\dagger (f(a))D(f(a))+R(f(a))
\end{equation}
as the shape-invariance condition. Typical solutions known until
now are as follows;
\begin{enumerate}
\item $f(a)=a-1, W(x,a)$ consists of finite power series of $a$.
In this case, we have six types of potentials classified by
Infeld and Hull\cite{IN}.
\item $f(a)=qa, W(x,a)=aW(ax)$. This corresponds to the self-similar
potentials discovered by Shabat\cite{SH} and Spiridonov\cite{SP}.
\end{enumerate}
It should be noted here that this interpretation of the self-similar
potentials makes it possible to apply an associated coherent state
in\cite{FU} to them.
We can consider other possibilities
except for the ansatz (4) which is known as
the simplest and the most typical shape-invariant potentials,
for example, the case where superpotentials have
many parameters $W_n=W(x,a_n,b_n,..)$\cite{CO},
or the case where there are many independent superpotentials
$W_{2n}=W_1(x,a_n), W_{2n+1}=W_2(x,a_n)$\cite{SK}, and so on.
See also\cite{LE} for the classification of various typical potentials
and recently discovered ones.

For shape-invariant potentials we can calculate the
eigenvalues in an algebraic way\cite{IN,GE}.
To see this, let us consider
the following sequence of hamiltonians
\begin{eqnarray}
& &H_0\equiv D^\dagger (a_0)D(a_0)+R(a_0)\nonumber\\
& &H_1\equiv D(a_0)D^\dagger (a_0)+R(a_0)=D^\dagger (a_1)D(a_1)
R(a_0)+R(a_1)\nonumber\\
& &\hskip5cm\vdots\nonumber\\
& &H_{n+1}\equiv D(a_n)D^\dagger (a_n)+\sum_{k=0}^{n}R(a_k)
=D^\dagger (a_{n+1})D(a_{n+1})+\sum_{k=0}^{n+1}R(a_k)\nonumber\\
& &\hskip5cm\vdots
\end{eqnarray}
For these hamiltonians we can see the following two properties
provided that there exist normalized 0-eigenvalue states
satisfying $D(a_n)|\psi_0(a_n)\rangle =0\quad (n=0,1,2,...)$:
i) $D^\dagger (a_n)D(a_n)$ and $D(a_n)D^\dagger (a_n)$ are superpartners
each other. Namely, they have the same eigenvalues except for the
0-eigenvalue of the former. ii) The lowest eigenvalue of $H_{n+1}$
is $\sum_{k=0}^{n+1}R(a_k)$ since $D^\dagger (a_{n+1})D(a_{n+1})$
has a 0-eigenvalue.
Combining these two properties, we can conclude that the $n$th
eigenvalue of $H_0$ and the corresponding eigenstate are given by
\begin{eqnarray}
& &E_n(a_0)=\sum_{k=0}^nR(a_k),\nonumber\\
& &|\psi_n(a_0)\rangle \propto D^\dagger (a_0)D^\dagger (a_1)\cdots
D^\dagger (a_{n-1})|\psi_0(a_n)\rangle .
\end{eqnarray}
Therefore, once we know the two important functions of $a$, i.e.,
$R(a)$ and $f(a)$, we can easily calculte eigenvalues and
eigenstates.

Note that the shape-invariance condition (6) resembles
that of the harmonic oscillator commutation relation.
It needs, however, extra reparametrization operation. This is the reason
we need different creation operators in Eq.(8)
as many as the bound states. To rewrite
Eq.(6) as a commutation relation and to express the excited states
in terms of the powers of a single creation operator,
we introduce the operator
$T$ defined by
\begin{equation}
T|\phi (x,a)\rangle =|\phi (x,f(a))\rangle ,
\end{equation}
which denotes, namely, the reparametrization of the parameter $a$.
Using this operator, we define
\begin{equation}
A_+(a)\equiv D^\dagger (a) T, \quad A_- (a)\equiv T^{-1}D(a) .
\end{equation}
Then we achieve the following expression of the shape-invariance
condition
\begin{equation}
[A_+(a), A_-(a)]=R(a).
\end{equation}
It should be noted that
this commutation relation is not closed in general, i.e.,
$R(a)$ is not commutative with $A_\pm$ though it is a constant.
It is, however, possible to construct a coherent state in the
following way.
Let us here assume that all $|\psi_0(a_n)\rangle \quad (n=0,1,...)$
are normalizable eigenstates, i.e., $H_0$ has infinite number of
bound states. Then after some calculation we get the expression of
the normalized eigenstate of Eq.(8)
\begin{equation}
|\psi_n(a_0)\rangle =\frac{1}{\sqrt{[n]_0!}}\{A_+(a_0)\}^n|\psi_0(a_0)
\rangle ,
\end{equation}
where
\begin{eqnarray}
& &[n]_k\equiv \sum_{i=1}^{n}R(a_{k+i}) ,\quad
\widehat{[n]}_k\equiv [n]_kT ,\nonumber\\
& &[n]_k!\equiv\widehat{[n]}_k\widehat{[n-1]}_k\cdots\widehat{[1]}_k
\cdot T^{-n} .
\end{eqnarray}
The appearance of $T$ in $\widehat{[n]}_k$ reflects the non-commutative
character between $R(a)$ and $A_\pm(a)$.

Now let us define a ``coherent state" associated with the commutation
relation Eq.(11). Here coherent state means the eigenstate of the
``annihilation" operator $A_-(a)$.
For this purpose, we first define the generalized exponential
function
\begin{equation}
\exp_k(x)\equiv\sum_{k=0}^\infty\frac{1}{[n]_k!}x^n ,
\end{equation}
using Eq.(13). Next, we define the state
\begin{eqnarray}
|z,a_0)&\equiv&\exp_0\{zA_+(a_0)\}|\psi_0(a_0)\rangle\nonumber\\
&=&\sum_{k=0}^\infty\frac{1}{\sqrt{[n]_0!}}z^n|\psi_n(a_0)\rangle .
\end{eqnarray}
{}From the direct calculation, we can easily confirm the relation
$A_-(a_0)|z,a_0)=z|z,a_0)$.
For further details, including the property of the completeness,
see\cite{FU}.

Now we concretely calculate the coherent state (15)
for the self-similar potentials. First, we must confirm that
there really appears the q-oscillator algebra,
since the shape-invariance
condition (11) is here represented by the usual oscillator-like
commutation relation.
By substitution $f(a)=qa$ and $W(x,a)=aW(ax)$ into Eq.(5), we have
the following difference-differential equation
\begin{equation}
W^2(x)+\frac{dW(x)}{dx}=q^2W^2(qx)-q\frac{dW(qx)}{dx}
+\frac{2R(f(a))}{a^2} .
\end{equation}
As previously mentioned, this equation should be satisfied identically
with respect to $a$. The last term should be, therefore, a constant,
denoted here by $\gamma (q) (>0)$:
\begin{equation}
R(a)=\frac{\gamma (q)}{2q^2}a^2 .
\end{equation}
Hereafter, we set $\gamma (q)=2$ for simplicity.
The commutation relation denoting the shape-invariance (11)
is then given by
\begin{equation}
[A_-(a), A_+(a)]=a^2/q^2 .
\end{equation}
What is important is that it is not closed, i.e.,
it needs infinite number
of generators to make it closed. For example, if we define
$A_-(a)\cdot a^n, a^n\cdot A_+(a)$ and $a^n$, where $n$ denotes integer,
we have some closed relation.
It is possible, however, to get a subalgebra
as follows: Let us introduce modified $A$-operators
\begin{eqnarray}
& &A_{q+}(a)\equiv \frac{1}{a}A_+(a) ,\nonumber\\
& &A_{q-}(a)\equiv A_-(a)\frac{1}{a}.
\end{eqnarray}
Then, the above commutataion relation is rewritten as
\begin{equation}
A_{q-}(a)A_{q+}(a)-q^2A_{q+}(a)A_{q-}(a)=1 ,
\end{equation}
which is essentially  equivalent to the one derived by
Spiridonov\cite{SP}.
This is quite natural since $T$ operator acts like a dilation
operator $T_qf(x)=f(qx)$ on $D_{q\pm}=(W(ax)\pm d/d(ax))/\sqrt{2}$,
where $A_{q\pm}\equiv D_{q+}T, T^{-1}D_{q-}$.

Next, let us calculate $[n]_0$ and $[n]_0!$ in Eq.(13).
By definition, we have $a_k=q^ka_0$ and therefore $R(a_k)=q^{2(k-1)}
a_0^2$. Then we have
\begin{eqnarray}
& &[n]_0=[n]_q\cdot a_0^2 ,\nonumber\\
& &[n]_0!=[n]_q!\cdot a_0^2a_1^2\cdots a_{n-1}^2 ,
\end{eqnarray}
where $[n]_q\equiv (1-q^{2n})/(1-q^2)$ is a q-deformed $n$ and
$[n]_q!\equiv [n]_q[n-1]_q\cdots [1]_q$ is a q-deformed factorial.

Before we calculate the coherent state (15), note that we define
this state based on the commutation relation (11).
What is important here is that this relation is invariant under
the transformation
\begin{eqnarray}
& &{\mathop{A}^\circ}_+(a)=g(a)A_+(a),\nonumber\\
& &{\mathop{A}^\circ}_-(a)=A_-(a)\frac{1}{g(a)},
\end{eqnarray}
where $g(a)$ is an arbitrary function of $a$. Using this property,
we can immediately define a coherent state which is the eigenstate
of ${\mathop{A}^\circ}_-$ as follows;
\begin{equation}
{\mathop{A}^\circ}_-|z,a_0)^\circ =z|z,a_0)^\circ
\end{equation}
with
\begin{equation}
|z,a_0)^\circ =\exp_0\{z{\mathop{A}^\circ}_+(a_0)\}
|\psi_0(a_0)\rangle .
\end{equation}
Now let us choose $g(a)=a$. Then we have
\begin{equation}
{\mathop{A}^\circ}_-(a)=A_{q-}(a), \quad
{\mathop{A}^\circ}_+(a)=a^2A_{q+}(a) .
\end{equation}
By the use of Eq.(21), the exponent in Eq.(24) is calculated as follows;
\begin{eqnarray}
\exp_0\{z{\mathop{A}^\circ}_+(a_0)\}&=&\sum_{n=0}^\infty
\frac{1}{[n]_q!\cdot a_0^2a_1^2\cdots a_{n-1}^2}
\{za_0^2A_{q+}(a_0)\}^n\nonumber\\
&=&\exp_q\{zA_{q+}(a_0)\} .
\end{eqnarray}
where $\exp_q(x)\equiv\sum_{n=0}^\infty x^n/[n]_q!$ is a q-deformed
exponential function.
Therefore, we conclude that a coherent state associated with
the shape-invariance naturally leads to the q-coherent state\cite{BI}
in case of the self-similar potentials.
The convergence radius of this $\exp_q(x)$ depends on q, especially,
in case of $|q|<1$, it is finite.
It is, however, possible to make it infinite if we construct
a coherent state by choosing, for example, $g(a)=a^2$ in Eq.(22).
The eigenvalue of $A_{q-}(a)$ for such a state is $(a/q)z$.

In summary, we have reformulated the self-similar potentials in
terms of the shape-invariance and calculated an associated coherent
state recently proposed in\cite{FU}.
We have found that it natually reduces to the q-coherent state
related to the q-oscillator.
Since the potentials of exactly solved many-body systems in one
dimension\cite{OL} are quite similar to the shape-invariant potentials,
it is interesting to generalize this idea to such many-body problems.
Especially, the exchange operator formalism recently developed
in\cite{PO} may have some connections with it.

The author would like to thank Dr. N. Aizawa for fruitful discussions.
This work is supported by the Grant-in-Aid from the Ministry of
Education, Science and Culture.


\begin{references}
\bibitem{KL} For a review of coherent states, see, e.g.,
J. R. Klauder, in {\it Coherent States.
Applicatons in Physics and Mathmatical Physics}, eds. J. R. Klauder
and B. S. Skagerstam (World Scientific, Singerpore, 1985).
\bibitem{PE} A. Perelomov,
{\it Generalized Coherent States and Their Applications}
(Springer-Verlag, Berlin, 1986).
\bibitem{FU} T. Fukui and N. Aizawa, to appear in Phys. Lett. {\bf A}.
\bibitem{IN} L. Infeld and T. E. Hull,
Rev. Mod. Phys. {\bf 23}, 21 (1951).
\bibitem{GE} L. E. Gendenstein, JETP Lett. {\bf 38} (1983) 356;
L. E. Gendenstein and I. V. Krive, Sov. Phys. Usp. {\bf 28}, 645 (1985).
\bibitem{SH} A. Shabat, Inver. Prob. {\bf 8}, 303 (1992).
\bibitem{SP} V. Spiridonov, Phys. Rev. Lett. {\bf 69}, 398 (1992).
\bibitem{SK} S. Skorik and V. Spiridonov, Lett. Math. Phys.
{\bf 28}, 59 (1993).
\bibitem{CO} F. Cooper, J. N. Ginocchio and A. Khare, Phys. Rev. {\bf
D36}, 2458 (1987).
\bibitem{LE} G. L\'evai, J. Phys. {\bf A22}, 689 (1989); J. Phys.
{\bf A25}, L521 (1992).
\bibitem{BI} L. C. Biedenharn, J. Phys. {\bf A22}, L873 (1989);
R. W. Gray and C. A. Nelson, J. Phys. {\bf A23}, L945 (1990);
A. J. Bracken, D. S. McAnally, R. B. Zhang and M. D. Gould, J. Phys.
{\bf A24}, 1379 (1991).
\bibitem{OL}  A. Olshanetsky and A. M. Perelomov,
Phys. Rep. {\bf 94}, 313 (1983), and references therein.
\bibitem{PO} A. P. Polychronakos, Phys. Rev. Lett. {\bf 69}, 703
(1992); L. Brink, T. H. Hansson and M. A. Vasiliev, Phys. Lett.
{\bf B286}, 109 (1992); L. Brink, T. H. Hansson, S. Konstein
and M. A. Vasiliev, Nucl. Phys. {\bf B401}, 591 (1993).

\end{references}
\end{document}